\documentclass{article}

\usepackage{arxiv}

\usepackage[utf8]{inputenc} % allow utf-8 input
\usepackage[T1]{fontenc}    % use 8-bit T1 fonts
\usepackage{hyperref}       % hyperlinks
\usepackage{url}            % simple URL typesetting
\usepackage{booktabs}       % professional-quality tables
\usepackage{amsfonts}       % blackboard math symbols
\usepackage{nicefrac}       % compact symbols for 1/2, etc.
\usepackage{microtype}      % microtypography
\usepackage{lipsum}
\usepackage{graphicx}
\graphicspath{ {./images/} }

\usepackage{amsmath,amssymb}
\usepackage{bm}
\usepackage{enumerate}
\usepackage{cite}
\newcommand{\mi}{\mathrm{i}}
\newcommand{\rr}{{\bf{r}}}

\title{Non-trivial retardation effects in dispersion forces:\\
From anomalous distance dependence to novel traps}

\author{
 Johannes~Fiedler\thanks{Centre for Materials Science and Nanotechnology, Department of Physics, University of Oslo, P. O. Box 1048 Blindern, NO-0316 Oslo, Norway}\\
Physikalisches Institut\\Albert-Ludwigs-Universit{\"a}t Freiburg\\Hermann-Herder-Str. 3\\79104 Freiburg, Germany.\\
\texttt{johannes.fiedler@physik.uni-freiburg.de}
   \And
   Kristian~Berland\\
Faculty of Science and Technology\\Norwegian University of Life Sciences, Campus {\AA}s\\Universitetstunet 3, NO-1430 {\AA}s, Norway.
\And
 Fabian~Spallek \\
  Physikalisches Institut\\Albert-Ludwigs-Universit{\"a}t Freiburg\\Hermann-Herder-Str. 3\\79104 Freiburg, Germany. \\
  \And
  Iver~Brevik\\
Department of Energy and Process Engineering\\Norwegian University of Science and Technology\\NO-7491 Trondheim, Norway. \\
\And
Clas~Persson\\Centre for Materials Science and Nanotechnology\\Department of Physics\\University of Oslo\\P. O. Box 1048 Blindern\\NO-0316 Oslo, Norway.\\
\And
Stefan~Yoshi~Buhmann\\
Physikalisches Institut\\Albert-Ludwigs-Universit{\"a}t Freiburg\\Hermann-Herder-Str. 3\\79104 Freiburg, Germany.
\And
Mathias~Bostr{\"o}m\footnotemark[1]\\%\thanks{Centre for Materials Science and Nanotechnology, Department of Physics, University of Oslo, P. O. Box 1048 Blindern, NO-0316 Oslo, Norway}\\
Department of Energy and Process Engineering\\Norwegian University of Science and Technology\\NO-7491 Trondheim, Norway.\\\texttt{mathias.a.bostrom@ntnu.no}
}

\begin{document}
\maketitle
\begin{abstract}
In the study of dispersion forces, nonretarded, retarded and thermal asymptotes with their distinct scaling laws are regarded as cornerstone results governing interactions at different separations. Here, we show that  when particles interact in a medium, the influence of retardation is qualitatively different, making it necessary to consider the non-monotonous potential in full. We discuss different regimes for several cases and find an anomalous behaviour  of the retarded asymptote. It can change  sign,  and lead to  a trapping potential.
\end{abstract}

% keywords can be removed

\section{Introduction}
Dispersion forces, which include the Casimir~\cite{Casi} and Lifshitz~\cite{Dzya} force between two dielectric bodies, the Casimir--Polder~\cite{Casimir48} force between a dielectric body and a polarisable particle, and the London--van der Waals force~\cite{London1930,vdw1873} between two polarisable particles, all arise from ground-state fluctuations of the electromagnetic fields. These forces which typically lead to  an attractive interaction between the constituents, have been much studied both experimentally~\cite{doi:10.1021/acsnano.7b05204,Schollkopf1994,Grisenti1999,Arndt1999} and and theoretically in great detail~\cite{Buhmann12a,Dalvit,Cherroret2015}. Systems in which dispersion forces act across a region of vacuum have received  most attention and a number of asymptotic results have been established for different geometries~\cite{PhysRevLett.104.070404}. In most cases a simple $r^{-n}$ distance dependence for the non-retarded and thermal limits (with integer $n$) and an $r^{-n-1}$ distance law connecting both asymptotes in the retarded regime is found. Related to this result is the fact that the proportionality constant of the non-retarded regime is larger than that from the thermal regime. When immersed in a medium~\cite{Ether_2015,LOSKILL2012107} the interaction between particles is screened. For these cases, the retarded asymptote is often not a useful approximation for the full theory, at least not so until it merges with the thermal asymptote. To add to the complex picture we show here that the full interaction curves for London--van der Waals, Casimir--Polder and Casimir--Lifshitz in a medium are not always monotonically decaying, and they potentially have more than one extreme point. They can under certain conditions become repulsive~\cite{PhysRevLett.106.064501,PhysRevA.81.062502}. Such repulsive forces can be balanced with other forces, such as buoyancy~\cite{doi:10.1021/acs.jpcc.8b02351,MathiasPhysRevB2018}, to stabilise a particle's position. Recently, a remarkable Lifshitz force induced trapping was experimentally observed exploiting a layered medium where short range repulsion was caused by a thin film coating, while larger distances were dominated by the underlying bulk material leading to attraction~\cite{Zhao984}. This effect was earlier theoretically predicted, e.g., by Dou \textit{et al.}~\cite{DouPhysRevB.89.201407}. Here, we demonstrate that a dispersion potential alone, even in the absence of a layered surface, has the ability to trap particles without the presence of any balancing forces. {In contrast to previous works, where for instance retardation effects of the Casimir force have been predicted~\cite{PhysRevA.85.064501} or measured~\cite{doi:10.1021/la981381l}, we demonstrate that the thermal limit is more important for medium-assisted dispersion forces. Furthermore, we demonstrate, the crossings of dielectric function also may yield  a breakdown of the retarded asymptote, leading to unexpected potential behaviours. 

\section{Medium-assisted Dispersion forces}
In the following, we analyse the different dispersion interactions in the presence of an environmental medium with respect to their asymptotic behaviours. 

\subsection{Medium-assisted London--van-der-Waals interactions}
The van der Waals interaction is the interaction between two neutral particles, $A$ and $B$. By applying perturbation theory to the atom--field Hamiltonian~\cite{PhysRevA.74.042101}, the energy change of the single systems can be obtained~\cite{Buhmann12a} (and references within)
\begin{eqnarray}
 \lefteqn{U_{\rm vdW}(\rr_A,\rr_B) = -\frac{\hbar\mu_0^2}{2\pi} \int\limits_0^\infty \mathrm d \xi \,\xi^4} \nonumber\\
 &&\times \operatorname{Tr} \left[\boldsymbol{\alpha}_A(\mi\xi)\cdot {\bf{G}}(\rr_A,\rr_B,\mi\xi)\cdot \boldsymbol{\alpha}_B(\mi\xi) \cdot {\bf{G}}(\rr_B,\rr_A,\mi\xi)\right]\,,\nonumber\\\label{eq:uvdwgen}
\end{eqnarray}
which can be interpreted (reading from right to left) as the propagation of a virtual photon, which is created at particle $A$, of frequency $\mi\xi$, that propagates to particle $B$ [${\bf{G}}(\rr_B,\rr_A)$], where it interacts with particle $B$ ($\boldsymbol{\alpha}_B$ and is send back to particle $A$ followed by the interaction with it. The sum (integral) over all these exchange processes results in the van der Waals potential. 

By inserting the bulk Green's function~\cite{Scheel2008}
\begin{eqnarray}
\lefteqn{{\bf{G}}(\rr,\rr',\omega) =  -\frac{1}{3k^2}\bm{\delta}({\bm{\varrho}})-\frac{\mathrm e^{\mi k\varrho}}{4\pi k^2\varrho^3}}\nonumber\\
&& \times\left\lbrace \left[1-\mi k\varrho - (k\varrho)^2\right] \mathbb{I} - \left[3-3\mi k\varrho -(k\varrho)^2\right]{\bf{e}}_{\varrho}{\bf{e}}_\varrho\right\rbrace,
\end{eqnarray}
with the relative coordinate $\bm{\varrho}=\varrho {\bf{e}}_\varrho$,
into Eq.~(\ref{eq:uvdwgen}) and applying the local-field corrections~\cite{Sambale07,Johannes,Friedrich}, the van der Waals potential between two  particles separated by a distance $\varrho$ embedded in a medium with permittivity $\varepsilon$ reads}
\begin{eqnarray}
 \lefteqn{U_{\rm vdW}(\varrho) =-\frac{\hbar}{16\varepsilon_0^2\pi^3\varrho^6}}\nonumber\\
 &&\qquad\times\int\limits_0^\infty\mathrm d \xi\,   \frac{\alpha_{\rm A}^\star(\mi\xi)\alpha_{\rm B}^\star(\mi\xi)}{\varepsilon^2(\mi\xi)} \,\,  g\left(\xi\varrho\sqrt{\varepsilon(\mi\xi)}/c \right)\, , \label{eq:vdW2}
\end{eqnarray}
with
\begin{equation}
    g(x) = \left(3+6x+5x^2 +2 x^3+x^4\right) \mathrm e^{-2x} \, .
\end{equation}
In Eq.~(\ref{eq:vdW2}), $\alpha_{\rm A,B}^\star$ are the environmentally screened polarisabilites. The three standard  asymptotic results for the van der Waals potential~\cite{PhysRevLett.104.070404} between two particles are as follows: 
\begin{enumerate}[i)]
    \item  The non-retarded regime, in which $\varrho n(0)$ (the optical path with refractive index at zero frequency) is significantly smaller than $c/\omega_{\rm max}$ with $\omega_{\rm max}$ the largest relevant transition frequency:
\begin{equation}
    U_{\rm vdW}^{\rm non-ret}(\varrho) =-\frac{3\hbar}{16\varepsilon_0^2\pi^3\varrho^6} \int\limits_0^\infty\mathrm d \xi\,   \frac{\alpha^\star_{\rm A}(\mi\xi)\alpha^\star_{\rm B}(\mi\xi)}{\varepsilon^2(\mi\xi)} \, ,
    \label{eq:non-ret}
\end{equation}
\item  the retarded regime, in
which  $n(0) \varrho\gg c/\omega_{\rm max}$: 
\begin{eqnarray}
   U_{\rm vdW}^{\rm ret}(\varrho) =-\frac{23\hbar c}{64\varepsilon_0^2\pi^3\varrho^7}\frac{\alpha^\star_{\rm A}(0)\alpha^\star_{\rm B}(0)}{\varepsilon^{5/2}(0)} \, ,
   \label{eq:ret}
\end{eqnarray}
\item  the thermal limit for separations larger than the thermal wave length $\varrho n(0) \gg \hbar c/(k_{\rm B} T)$, which leads to
\begin{equation}
     U_{\rm vdW}^{\rm th}(\varrho)  = -\frac{3k_{\rm B} T}{16\varepsilon_0^2\pi^2\varrho^6}    \frac{\alpha^\star_{\rm A}(0)\alpha^\star_{\rm B}(0)}{\varepsilon^2(0)}\,. \label{eq:thermal}
\end{equation}
\end{enumerate}

\begin{figure}[t]
    \centering
    \includegraphics[width=0.7\columnwidth]{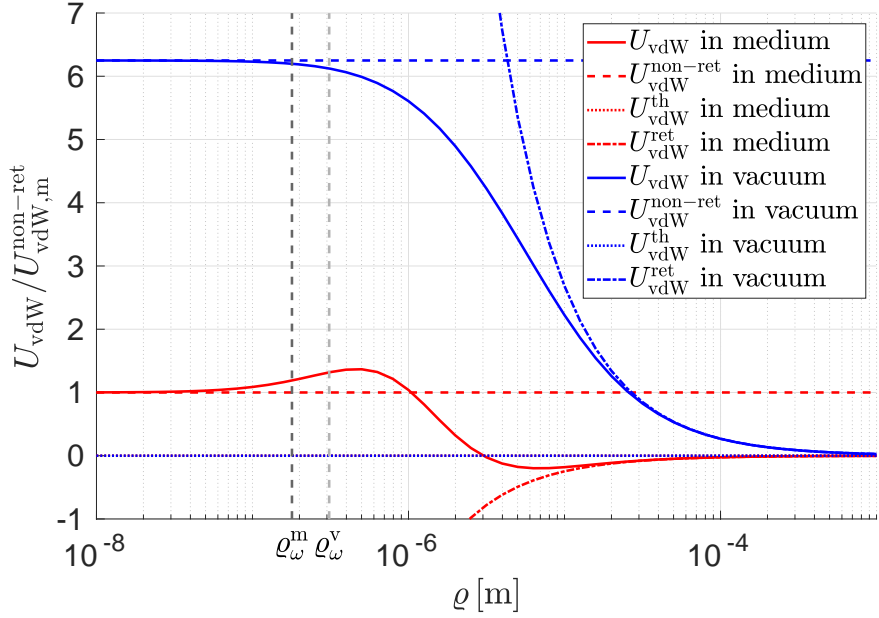}
    \caption{Van der Waals potential normalised to the medium-assisted nonretarded limit and the corresponding asymptotes (dashed lines the non-retarded limit, dashed-dotted the retarded limit and  dotted lines lines the thermal limit) for a vacuum scenario (blue lines) and the corresponding case embedded in a medium (red lines).}
    \label{fig:sample}
\end{figure}

To illustrate how immersion in a medium (M) can affect these power laws, we consider the interaction between two spherical nanoparticles of radius $a_{\rm A}$ and $a_{\rm B}$  with response functions given by single-oscillator model dielectric functions  $\varepsilon_{i}(\mi\xi)=1 +\chi_{ i}^{(0)}/[1+(\xi/\omega_{ i})^2]$ with $i = {\rm A, B, M}$ with amplitudes $\chi_{\rm A}^{(0)}=1$, $\chi_{\rm B}^{(0)}=4$, $\chi_{\rm M}^{(0)}=2$  and resonance frequencies $\omega_{\rm A}=4\,\rm{eV}$, $\omega_{\rm B}= (1/4) \,\rm{eV}$  and $\omega_{\rm M}= (1/2) \,\rm{eV}$. In this model, damping effects are neglected because its impact does not influence the potential qualitatively. The corresponding polarisabilities are computed via the Clausius--Mossotti relation~\cite{Jackson} for interactions through a void
\begin{equation}
 \alpha_{\rm vac}(\omega) = 4\pi a^3 \varepsilon_0 \frac{\varepsilon(\omega)-1}{\varepsilon(\omega) +2} \, ,
\end{equation}
and the hard-sphere model for interactions through a medium~\cite{Johannes}
\begin{equation}
 \alpha_{\rm HS}(\omega) = 4\pi a^3 \varepsilon_0\varepsilon_{\rm M}(\omega) \frac{\varepsilon(\omega)-\varepsilon_{\rm M}(\omega)}{\varepsilon(\omega) +2\varepsilon_{\rm M}(\omega)} \, ,
\end{equation}
with $a$ being the particle radius.

\begin{figure}[t]
    \centering
    \includegraphics[width=0.7\columnwidth]{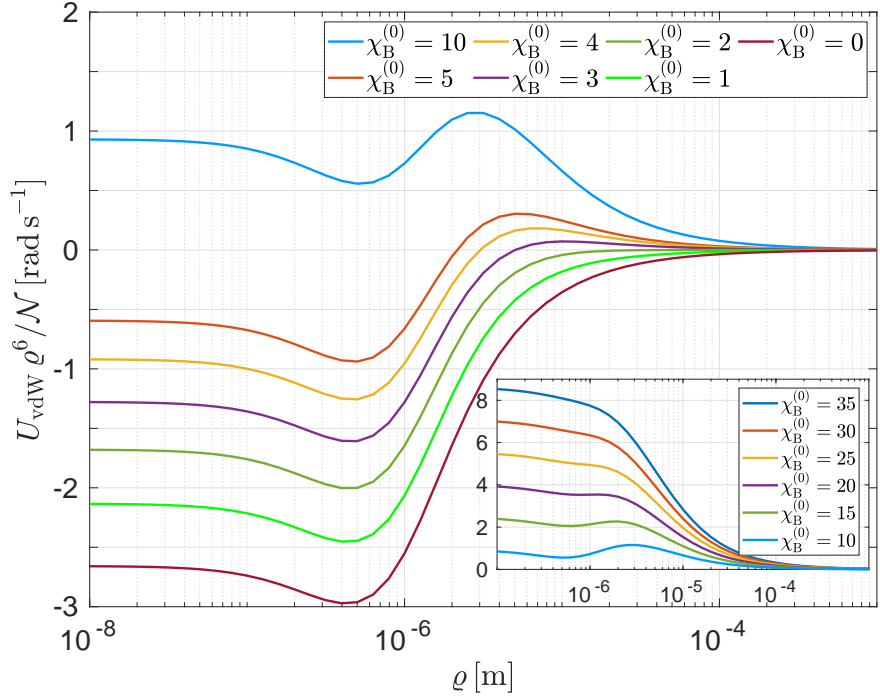}
    \caption{Dependence of the van der Waals potential by changing the dielectric's amplitude of one material with the normalisation factor $\mathcal{N} = 10^{13}\hbar a_{\rm A}^3a_{\rm B}^3/\pi$. The resonances stay fixed.}
    \label{fig:sample2}
\end{figure}

Figure~\ref{fig:sample} shows the resulting van der Waals potentials for both the vacuum  and the medium-assisted cases compared with the asymptotic expressions of Eqs.~(\ref{eq:non-ret})--(\ref{eq:thermal}). For clarity, the vertical axis is scaled by the sixth power of $\varrho$ ($U_{\rm vdW,m}^{\rm non-ret} = -C_6^{\rm m}/\varrho^6$). In the vacuum case (blue curves), we recover the expected limits: the exact potential follows the non-retarded asymptote for small separations ($\varrho<\varrho_\omega^{\rm v}= c/\omega_{\rm max}$); thereafter, it follows the retarded asymptotic form, until it finally matches the thermal limit whose amplitude is smaller than that of the nonretarded limit due to the steeper decrease in the retarded regime. This result is in agreement with the literature~\cite{doi:10.1021/la981381l}. Depending on the optical response functions such, similar results can also be found in the medium-assisted case. However, if there are crossings of the dielectric functions the behaviour of the asymptotes dramatically changes. The results for the medium-assisted case (red curves) differ drastically from these vacuum findings. First of all, retardation effects kick in at significantly shorter separations, which can be attributed to the refractive index of the medium $\varrho_\omega^{\rm m}=c/[\omega_{\rm max} n(0)]$. However, it is in the intermediate regime that a truly surprising behaviour emerges. In this regime, the potential exhibits additionally distinct extreme points in the retarded regime. In fact, for this example, the retarded asymptote does not become a viable approximation for the full theory until it merges with the thermal asymptote. There is hence in effect no regime at all for which the retarded asymptote, Eq.~(\ref{eq:ret}) gives a good description. Finally, the prefactor of the nonretarded and thermal asymptotes differ in sign. Such potential minima have been found in earlier studies, for instance in connection with surface wetting~\cite{Elbaum273Kwater}; however, here we relate this to the breakdown of the retarded regime. In Fig.~\ref{fig:sample2}, we explore how adjusting the amplitude of the dielectric function of particle B affects the shape of the potential with all other parameters kept fixed, as it influences the number of intersections with the dielectric functions of the medium and particles ($\chi^{(0)}_{\rm B} = 0,1,\dots,10,15,20,25,30,35$). It can be observed that  as $\chi^{(0)}_{\rm B}$ increase, the potential changes from attraction to repulsion at short distances. For amplitudes $\chi^{(0)}_{\rm B}<2$ lower than that of the medium, only one minimum can be found according to the crossing of dielectric function of the medium and particle. For higher values a second extreme point at at larger separation occurs due to the crossing of the dielectric functions and the lower resonant frequency of particle ($\omega_{\rm B}$) compared to that of the medium ($\omega_{\rm M}$). For very large amplitudes $\chi^{(0)}_{\rm B}>20$, the effect of the crossings vanishes and the potential becomes smoother.

\subsection{Medium-assisted Casimir--Polder interaction}
The anomalous asymptotic behaviour is not restricted to particle--particle interactions. For similar crossings of dielectric functions, it can also be present for a particle with polarisability $\alpha$  in front of a solid plate with permittivity $\varepsilon_{\rm M}(\mi\xi)$ at temperature $T$, immersed in a liquid medium  with permittivity $\varepsilon_{\rm L}(\mi\xi)$.

{The Casimir--Polder interaction, in general, describes the force between a neutral particle with polarisability $\boldsymbol{\alpha}$ in the presence of dielectric objects with permittivity $\varepsilon(\rr,\omega)$. In analogy to the van der Waals potential, the Casimir--Polder energy can be derived via the application of perturbation theory to the atom-field Hamiltonian with a single particle~\cite{Buhmann_2004}. The result can be written as~\cite{Scheel2008}
\begin{equation}
    U_{\rm CP}(\rr) =\frac{\hbar\mu_0}{2\pi}\int\limits_0^\infty \mathrm d \xi \, \xi^2 \operatorname{Tr}\left[\boldsymbol{\alpha}(\mi\xi) \cdot {\bf{G}}(\rr,\rr,\mi\xi)\right]\,.
\end{equation}
By using the Green's function for planarly layered media~\cite{PhysRevA.51.2545},
\begin{eqnarray}
{\bf{G}}(\rr,\rr',\omega) &=& \frac{\mi}{8\pi^2}\int\frac{\mathrm d^2 k^\parallel}{k_1^\perp}\mathrm e^{\mi {\bf{k}}^\parallel \cdot \left(\rr-\rr'\right) + \mi k_1^\perp \left(z+z'\right)}\nonumber\\
 &&\times \sum_{\sigma=s,p} r_\sigma^{12}{\bf{e}}_{\sigma+}^1 {\bf{e}}_{\sigma-}^1 \, ,
\end{eqnarray}
with the reflection coefficients
\begin{eqnarray}
 r_s^{12}= \frac{k_1^\perp-k_2^\perp}{k_1^\perp+k_2^\perp}\,,\quad
  r_p^{12}= \frac{\varepsilon_2 k_1^\perp-\varepsilon_1 k_2^\perp}{\varepsilon_2 k_1^\perp+ \varepsilon_1 k_2^\perp}\,,
\end{eqnarray}
again applying the local-field corrections for the excess polarisability, and introducing the temperature dependence according to the standard substitution for the integral to the Matsubara sum
\begin{eqnarray}
   \lefteqn{ 2\pi\hbar \int \mathrm d\xi f(\xi)\mapsto k_{\rm B} T \sum_{n=0}^\infty{}' f(\xi_n)}\nonumber\\
   &&=k_{\rm B} T\left[\frac{1}{2}f(\xi_0)+ \sum_{n=1}^\infty f(\xi_n)\right] \, ,
\end{eqnarray}
with respect to the non-resonant part of the Casimir--Polder potential~\cite{Buhmann12b}, with the primed sum denotes a sum over Matsubara frequencies $\xi_n = 2\pi n k_{\rm B} T/\hbar$, with the first term weighed by 1/2~\cite{Buhmann12a},} the Casimir--Polder potential is given by~\cite{Buhmann12a} \begin{eqnarray}
 \lefteqn{U_{\rm CP} (z_{\rm A}) = \frac{k_{\rm B} T \mu_0}{4\pi} \sum_{n=0}^\infty {}' \alpha^\star(\mi\xi_n) \xi_n^2} \nonumber\\
 &&\times\int\limits_{0}^\infty \mathrm d k \frac{k}{\kappa_{\rm L}^\perp} \mathrm e^{-2\kappa_{\rm L}^\perp z_{\rm A} } \left[\frac{\kappa_{\rm L}^\perp - \kappa_{\rm M}^\perp}{\kappa_{\rm L}^\perp+\kappa_{\rm M}^\perp}\right.\nonumber\\
 &&\left.- \left(1+2\frac{{\kappa_{\rm L}^\perp}^2c^2}{\xi_n^2\varepsilon_{\rm L}(\mi\xi)}\right)\frac{\varepsilon_{\rm M}\kappa_{\rm L}^\perp -\varepsilon_{\rm L} \kappa_{\rm M}^\perp}{\varepsilon_{\rm M}\kappa_{\rm L}^\perp +\varepsilon_{\rm L} \kappa_{\rm M}^\perp} \right] \,,
\end{eqnarray}
with the imaginary part of the perpendicular wave vectors given by $\kappa_i^\perp=\sqrt{\varepsilon_i(\mi\xi)\xi^2/c^2+k^2}$. It takes the following forms in the non-retarded asymptote
\begin{equation}
    U_{\rm CP}^{\rm non-ret}(z) = - \frac{C_3}{z^3}  \, ,\label{eq:UCP}
\end{equation}
with
\begin{equation}
    C_3 =  \frac{k_{\rm B}T}{8\pi\varepsilon_0} \sum_{n=0}^\infty {}' \frac{\alpha^\star(\mi\xi_n)}{\varepsilon_{\rm L}(\mi\xi_n)} \frac{\varepsilon_{\rm M}(\mi\xi_n)-\varepsilon_{\rm L}(\mi\xi_n)}{\varepsilon_{\rm M}(\mi\xi_n)+\varepsilon_{\rm L}(\mi\xi_n)} \, ,\label{eq:C3}
\end{equation}
and the retarded asymptote 
\begin{equation}
    U_{\rm CP}^{\rm ret}(z) = -\frac{C_4}{z^4} \,,
\end{equation}
with
\begin{eqnarray}
 \lefteqn{ C_4= \frac{3\hbar c\alpha^\star(0)}{64\pi^2 \varepsilon_0 \varepsilon_{\rm L}^{3/2}(0)}\int\limits_1^\infty \mathrm d v}\nonumber\\
 &&\times  \, \left[\left(\frac{2}{v^2} -\frac{1}{v^4}\right)\frac{\varepsilon_{\rm M}\sqrt{\varepsilon_{\rm L}}v -\varepsilon_{\rm L}\sqrt{\varepsilon_{\rm L}(v^2-1)+\varepsilon_{\rm M}}}{\varepsilon_{\rm M}\sqrt{\varepsilon_{\rm L}}v +\varepsilon_{\rm L}\sqrt{\varepsilon_{\rm L}(v^2-1)+\varepsilon_{\rm M}}}\right.\nonumber\\
 &&\left.-\frac{1}{v^4}\frac{\sqrt{\varepsilon_{\rm L}}v-\sqrt{\varepsilon_{\rm L}(v^2-1)+\varepsilon_{\rm M}}}{\sqrt{\varepsilon_{\rm L}}v+\sqrt{\varepsilon_{\rm L}(v^2-1)+\varepsilon_{\rm M}}}\right] \, ,\label{eq:C4}
\end{eqnarray}
and finally, in the thermal limit, it is given by
\begin{equation}
    U_{\rm CP}^{\rm th}(z) = - \frac{C_{\rm 3T}}{z^3} \, ,
\end{equation}
with
\begin{equation}
    C_{\rm 3T} = - \frac{k_{\rm B}T}{16\pi\varepsilon_0} \frac{\alpha^\star(0)}{\varepsilon_{\rm L}(0)} \frac{\varepsilon_{\rm M}(0)-\varepsilon_{\rm L}(0)}{\varepsilon_{\rm M}(0)+\varepsilon_{\rm L}(0)} \, . \label{eq:C3T}
\end{equation}
The behaviour of the three asymptotes in vacuum is similar to the previously discussed van der Waals case depicted in  Fig.~\ref{fig:sample} with modified power laws ($r^{-3}$ for nonretarded, $r^{-4}$ for retarded and $r^{-3}$ for the thermal asymptote). Further details can be found in Ref.~\cite{PhysRevA.85.042513}. 

\begin{figure}
\centering
\includegraphics[width=0.7\columnwidth]{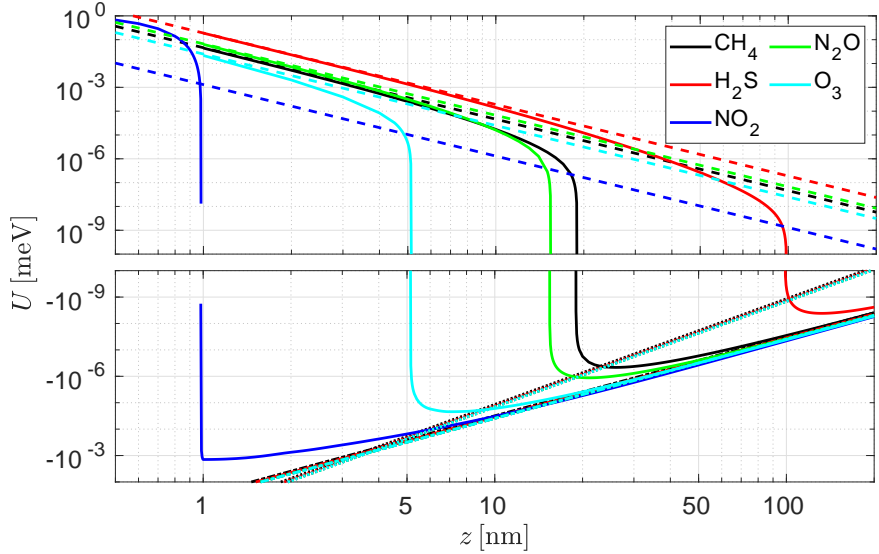}
\caption{The Casimir--Polder potential for dissolved CH$_4$ molecule (black), H$_2$S (red), NO$_2$ (blue), N$_2$O (green) and O$_3$ (cyan) near water surface shown together with the two limiting asymptotes (non-retarded dashed), and retarded (dotted).}
\label{Fig44}
\end{figure}

Applying this model to a specific example, we consider greenhouse gases with polarisabilities and particle and cavity radii taken from Ref.~\cite{Johannes}. The gases immersed in water interact with the water--air interface. Further, we use the finite-size model of Refs.~\cite{Sambale07,Johannes} to describe the excess polarisability that arises when taking into account the finite-size effects of the particles and a vacuum layer surrounding the particle to  avoid the contact between particle and solvent~\cite{Held2014}. The calculated potentials and the corresponding asymptotes are depicted in Fig.~\ref{Fig44}. The corresponding parameters for the asympotes, traping distances and frequencies are given in Table~\ref{tab:C3-}. It can be observed that hydrogen sulfide ($z_{\rm{trap}}=132\,\rm{ nm}$), methane (26 nm),  nitrogen dioxide (21 nm), ozone (7 nm) and nitrous oxide (1 nm) are trapped near the surface. For these systems, a potential minimum occurs between the non-retarded and thermal limits and the retarded limit does not appear as an $r^{-4}$ power law. Some correlation between the trapping distance $z_{\rm trap}$ and the position corresponding to the change in sign of dielectric function $z_0=c/[\omega_0 n(0)]$ with $\alpha(\mi\omega_0)=0$ can be found, as well as between the trapping distance and the transition distance between the retarded and non-retarded limit $z_\omega$. However, a simple correlation between one of them alone and the trapping distances cannot be expected, because the potential depends on all three quantities: the dielectric functions of the medium and the polarisabilities of the particles, whereas each of these parameters only depends on one of them. The gas molecule H$_2$S deviates from most other gas molecules in the non-retarded limit in having a very large negative C$_3$ in Table~\ref{tab:C3-}.  One would have expected H$_2$S to be hydrophobic as was recently discussed in Ref.~\cite{FiedlerWaterJPCB2020}. This result will require further investigation.

\begin{table}[tb]
    \centering
    \begin{tabular}{cccccccc}\hline
        Mol. & C$_3$ & C$_4$ & C$_{\rm 3T}$ & z$_{\rm trap}$ &  $z_0$ & $z_\omega$ & $\omega_{\rm trap}$\\\hline
         CH$_4$ &  -45.9 &11638  &31.3 & 26.0  &  59.2 &1.0 & 6.5\\
         CO     &  109.9  &12755  &34.3 & -     &  0.8  & 0.2& -\\
         CO$_2$ & 31.3   &15464  &41.6 &  -    &  2.4  &0.3 & -\\
         H$_2$S &  -189.9 &12958  &34.8 & 132.1 & 142.1 &2.2 & 0.1\\
         N$_2$  &  66.4   &11181  &30.1 & -     &  1.5  & 0.3& -\\
         NO$_2$ & -1.3    &14466  &38.9 & 1.0   &  3.6  &0.3 & 3930 \\
         N$_2$O & -66.8   &14574  &39.2 & 21.0  &  41.4 & 0.5& 7.8\\
         O$_2$  & 96.4    &10903  &29.3 & -     &  1.1  &0.2 & -\\
         O$_3$  &-24.9    &14562  &39.2 & 7.2   &  26.0 &0.3 & 90.8\\\hline
    \end{tabular}
    \caption{Data of $C_3$-coefficients ($[\mu\mathrm{eV (nm)^3}]$, eq.~(\ref{eq:C3})), 
    $C_4$-coefficients ($[\mu\mathrm{eV (nm)^4}]$, eq.~(\ref{eq:C4})), $C_{\rm 3T}$-coefficients ($[\mu\mathrm{eV (nm)^3}]$, eq.~(\ref{eq:C3T})), trapping distances (z$_{\rm trap}$ [nm]), and the corresponding trapping frequency ($\omega_{\rm trap}$ [MHz]) for different molecules dissolved in water at $T=273.16\,\rm{K}$ near a water--air surface. Positive $C_3$, $C_4$, and $C_{\rm 3T}$ values correspond to attraction. Further, the distances for the impact of retardation ($z_\omega$ [nm]) are given for each case. The thermal effects occur on distances larger than $z_{\rm T}=893$\,nm. As an approximation of the trapping distance tabled the values for $z_0$ that corresponds to the roots of the polarisabilities.}
    \label{tab:C3-}
\end{table}

By comparing the depth of the potential minimum with the thermal energy one finds that the stability of the trap is given at temperatures far below the freezing temperature of water. This means that for these examples, the interactions are not strong enough to trap particles. In terms of particles dissolved in a medium, one might be able to discern a slightly higher concentration of particles at these specific trapping distances. We note that the thermal stability of the trap increases with decreasing trap distance. Further, the energetic minimum of the potential  becomes steeper with decreasing trap distance resulting in a narrowed trapping potential the closer it gets to the interface. For  different materials these trends will be similar. 

Single gas molecules reveal anomalous interactions near a water surface, but the typical energy minimum is not sufficiently deep to act as an effective trap. Inspired by the observed Casimir--Polder repulsion for air bubbles in water near solid surfaces~\cite{PhysRevLett.106.064501,Esteso2019Langmuir} we consider a system where we expect trapping could occur. Specifically, we consider an air bubble dissolved in liquid bromobenzene~\cite{PhysRevA.81.062502} in front of an  horizontal, or vertical, amorphous silica surface. For modelling, we used parameters corresponding to amorphous silica (Volume of the SiO$_2$ unit: $V_{\rm v}$=41.14\AA$^3$~\cite{C5CP06775H}) at room temperature ($T=298\,\rm{K}$). The air bubble can, due to a crossing of dielectric functions for amorphous silica and bromobenzene at a specific frequency, be trapped, via short range repulsive and long range attractive Casimir--Lifshitz forces~\cite{Buhmann12a}. Here, we apply a simple version of the Derjaguin approximation to estimate the force on the sphere of radius $R$, $f_{\rm sphere}(x)=2 \pi R U_{\rm plane}(x)$ with  $U_{\rm plane}(x)$ being the energy for planar system. We then integrate the force from infinity up to the specific distance $x$, to obtain the interaction free energy acting on a sphere at a distance $x$ from the planar interface. The minimum sphere radius $R_{\rm min}$ will be estimated from the size that provides a trapping energy larger or equal to $k_{\rm B} T$. Here, we find an estimate for a thermally stable position of approximately {7\,nm} in front of the surface for bubbles with a radius much larger than $R_{\rm min}$=200\,nm. Hence, we are in a range for relative sizes and distances where the Derjaguin approximation is appropriate to use, but where surface roughness and various other nanoscale effects are important. The short range repulsion experienced by the typical air bubble is expected to lead to low surface friction. We note that for bubbles, below or above a surface, there will also be some influence on trapping distance from buoyancy, $b=V g  (\rho_{\rm l}-\rho_{\rm a})$, which depends on volume ($V$), gravitational constant $g$ and the difference in mass density for bubble $\rho_{\rm a}$ and the liquid $\rho_{\rm l}$. However, in front of a vertical surface buoyancy only acts to move the air bubble upwards while the proposed mechanism keeps it moving with low friction at a fixed trapping distance from the surface.

\section{Application: tuneable trap}

Ideally, a larger trapping distance than found above would be more realistic for experimental realisation. Thus, to introduce a scalable parameter for tuning the trap, we consider in our final example a two-component fluid surrounding the particle. For the dielectric functions of mixtures between fluid 1 ($\varepsilon_1$) in fluid 2 ($\varepsilon_2$)  we use a Lorentz--Lorenz model~\cite{Aspnes,Bostrom_acsearthspacechem.9b00019,MathiasPhysRevB2018}, where we introduce the volume fraction $p$ of fluid 1 in fluid 2. We chose the liquids bromobenzene and methanol in front of a polystyrene surface~\cite{PhysRevA.81.062502} as the dielectric function of the latter lies between both fluids. An illustration of the resulting dielectric functions can be found in Ref.~\cite{physrevapplied}. In this case, the crossings of the dielectric functions depend on the volume fraction.
\begin{figure}[t]
\centering
 \includegraphics[width=0.7\columnwidth]{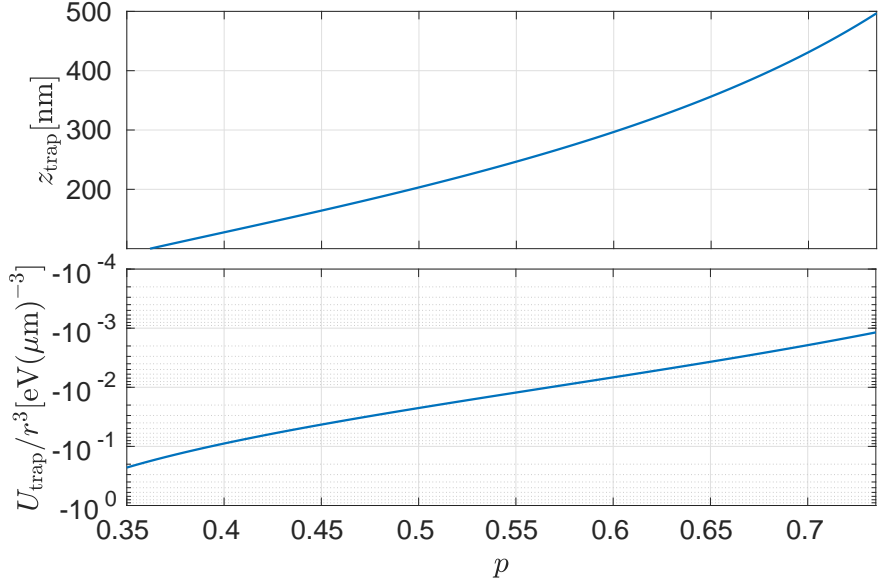}
 \caption{Trap parameters depending on the volume fraction $p$.}\label{fig:para}
\end{figure}
We used the example of an anatase-TiO$_2$ spherical nanoparticle whose dielectric function was taken from Ref.~\cite{doi:10.1063/1.4793273}. As we consider larger particles compared to few-atomic molecules from the example above, we describe the excess polarisability via the hard-sphere model as in the introductory van der Waals example. The resulting parameters for the trap are given in Fig.~\ref{fig:para}. We predict that the trapping can be tuned over a wide range of distances (100-500\,nm) by changing the liquid mixture. 

\section{Results and Conclusion}

To conclude, we have shown that the impact of retardation dramatically changes the asymptotic behaviour of dispersion interactions in media. In contrast to the ordinary theory in vacuum, the retarded power laws are not applicable and a consideration of the full interaction potential is required instead. Further, the transition distance between non-retarded and retarded regimes strongly decreases by immersing the interacting particles in a  liquid due to its refractive index. This extends previous work on Casimir--Lifshitz forces in fluids (see e.g. Refs.~\cite{Dzya,Elbaum273Kwater,PhysRevA.78.032109,PhysRevA.81.062502,MathiasPhysRevB2018,Zhao984,doi:10.1021/la981381l}) to a more general case where retarded dispersion forces can reveal a very complex behaviour in media. For this reason, considerations of retardation effects are important for medium-assisted dispersion force experiments, for example Casimir experiments~\cite{PhysRevLett.106.064501,PhysRevA.78.032109,doi:10.1021/la981381l}; medium-assisted optical tweezers~\cite{Gao2017}; colloidal systems~\cite{Brugger2015}; and in future measurements of the Casimir torque in a medium~\cite{PhysRevLett.119.183001,PhysRevB.100.205403}. Beyond these impacts, we have shown that the near-field effect of dispersion forces in colloidal system yields an inhomogeneous particle density distribution near interfaces due to trapping potentials. The introduced mechanism can for instance be used to trap nanoparticles at low temperature, specifically nanodiamonds~\cite{Dumeige:19}, by choosing liquid nitrogen, liquid fluorine, or atomic clouds as an environmental medium. The presented theory is more general and can be applied to several systems beyond Casimir experiments, especially medium-assisted spectroscopy in nanodroplets~\cite{doi:10.1063/1.4759443,doi:10.1021/acs.jpca.8b06673,doi:10.1063/1.5097553}. 

\section*{Acknowledgments}
We acknowledge support from the Research Council of Norway (Project  250346). We gratefully acknowledge support by the German Research Council (grant BU 1803/6-1, S.Y.B. and J.F., BU 1803/3-1, S.Y.B. and F.S.).

\bibliographystyle{unsrt}  
\bibliography{bibi3.bib}  %%% Remove comment to use the external .bib file (using bibtex).

%\begin{thebibliography}{1}

%\bibitem{kour2014real}
%George Kour and Raid Saabne.
%\newblock Real-time segmentation of on-line handwritten arabic script.
%\newblock In {\em Frontiers in Handwriting Recognition (ICFHR), 2014 14th
%  International Conference on}, pages 417--422. IEEE, 2014.

%\bibitem{kour2014fast}
%George Kour and Raid Saabne.
%\newblock Fast classification of handwritten on-line arabic characters.
%\newblock In {\em Soft Computing and Pattern Recognition (SoCPaR), 2014 6th
%  International Conference of}, pages 312--318. IEEE, 2014.

%\bibitem{hadash2018estimate}
%Guy Hadash, Einat Kermany, Boaz Carmeli, Ofer Lavi, George Kour, and Alon
%  Jacovi.
%\newblock Estimate and replace: A novel approach to integrating deep neural
%  networks with existing applications.
%\newblock {\em arXiv preprint arXiv:1804.09028}, 2018.

%\end{thebibliography}

\end{document}